\documentstyle[amssymb,12pt,epsfig]{article}
\begin{document}
\baselineskip=22pt

\hfill{\bf BIHEP-TH-2004-1}

\hfill{\bf USTC-ICTS-04-02}

\begin{center}
{\Large \bf Cosmic ray threshold anomaly \\ and kinematics in the
dS spacetime
}\\
\vspace{1cm}
Zhe Chang\footnote{changz@mail.ihep.ac.cn}~~and~~Shao-Xia Chen\footnote{ruxanna@mail.ihep.ac.cn}\\
{\em Institute of High Energy Physics,
Chinese Academy of Sciences} \\
{\em P.O.Box 918(4), 100039 Beijing, China}\\
Cheng-Bo Guan\footnote{guancb@ustc.edu.cn}\\
{\em Interdisciplinary Center for Theoretical Study}\\
{\em University of Science and Technology of China, 230026 Hefei, China}\\
\end{center}
\vspace{1.5cm}
\begin{abstract}
We present a covariant framework of kinematics in the dS
spacetime, which is a natural postulate of recent astronomical
observations ($\Lambda>0)$. One-particle states are presented
explicitly. It is noticed that the dispersion relation of free
particles is  dependent on the degrees of freedom of angular
momentum and spin. This fact can be referred to as the effects of
the cosmological constant on kinematics of particles. The
kinematics in dS spacetime is used to investigate the phenomenon
of ultra high energy cosmic rays.  We emphasize the possibility of
solving the threshold anomalies of the interactions between ultra
high energy cosmic rays and soft photons in the covariant
framework of kinematics in the dS spacetime.
 \vspace{1.0cm}
\end{abstract}
\newpage
\section{Introduction}

The origin of the ultra-high energy cosmic rays (UHECR) is one of
the outstanding puzzles of modern astrophysics. Today's
understanding of the phenomena responsible for the production of
UHECR is still limited. Currently, there are generally two
categories of production mechanisms of the UHECR. One is the
``bottom-up" acceleration scenario with some astrophysical objects
as sources\cite{acc1,acc2}. The other is called ``top-down"
scenario in which UHECR particles are from the decay of certain
sufficiently massive particles originating in the early
Universe\cite{top1}.

Decades ago, Greisen, Zatsepin and Kuzmin\cite{GZK} discussed the
propagation of the UHECR particles through the cosmic microwave
background radiation (CMBR). Due to the photopion production
process by the CMBR, the UHECR particles will lose their energies
drastically down to a theoretical threshold, which is about
$5\times 10^{19}$eV. That is to say, the mean free path for this
process is only a few Mpc\cite{GZK1}. This is the so-called GZK
cutoff. However, we have observed indeed hundreds of events with
energies above $10^{19}$eV and about 20 events above
$10^{20}$eV\cite{data1}-\cite{data5}. At the same time, there is
another paradox\cite{data6} in the terrain of cosmic ray, which
comes from the detected $20$TeV photons from the MrK $501$ (a BL
Lac object at a distance of $150$Mpc). Similar to the case of
UHECR, due to pair production process by the IR background
photons, the $20$TeV photons should have disappeared before
arrival at the ground-based detectors. Both of the puzzles can be
considered to be some cosmic ray threshold anomalies: energy of an
expected threshold is reached but the threshold has not been
observed yet.

Recently, there is growing interest on the theory of doubly
special relativity (DSR)\cite{DSR1, DSR2}. In the DSR, there exist
two observer-independent scales. One of them is a scale of
velocity, which is identified with the velocity of light. The
other is a scale of mass/length, which is expected to be the order
of the Planck mass/length. In fact, the violation of the Lorentz
invariance and the Planck scale physics have long been studied as
possible solutions of the cosmic ray threshold
anomalies\cite{loren2}-\cite{loren8}. However, all of these
scenarios are far beyond the standard cosmological theory and the
standard model of particle physics. Furthermore, it is well-known
that we are still very far from a theory of quantum gravity, in
spite of extensive investigations on candidates such as the
supergravity, Kluza-Klein, noncommutative geometry and superstring
theory.

The recent astronomical observations on supernovae
\cite{constant-super-1, constant-super-2} and CMBR
\cite{costant-cmbr} show that about two thirds of the whole energy
in the Universe is contributed by a small positive cosmological
constant $(\Lambda)$. An asymptotic de Sitter (dS) spacetime is
premised naturally. The physics in an asymptotic dS spacetime has
been discussed extensively\cite{ds-paper-1}-\cite{ds-paper-3}.

In this paper, we discuss kinematics in an asymptotic dS
spacetime. The framework of classical as well as quantum
kinematics in the dS spacetime is set up carefully. We get a
general form of dispersion relation for free particles in the dS
spacetime. This formalism is used to describe the UHECR
propagating in the cosmic microwave background as well as the
TeV-$\gamma$ propagating in the infrared background. We obtain
explicitly the corrections of the GZK threshold for the UHECR
particles interacting with soft photons, which  are dependent on
the cosmological constant as supposed in the beginning of the
paper. We show how the threshold varies with a positive
cosmological constant and additional degrees of freedom of the
angular momentums of interacting particles. It should be noticed
that, for a positive cosmological constant, the theoretic
threshold tends to be above the energies of all the observed
events. Thus, we may conclude that the tiny but nonzero
cosmological constant is a possible origin of the threshold
anomalies of the UHECR and the TeV-$\gamma$.

The paper is organized as follows. In Section 2, we discuss the
classical kinematics in the dS spacetime. Conservation laws of
momentum and angular momentum are obtained along the geodesics.
Section 3 is devoted to the investigation of the quantum
kinematics in the dS spacetime. By solving equations of motion of
a free particle, we present a remarkable dispersion relation for
free particles in the dS spacetime, which includes degrees of
freedom of angular momentum and spin. In Section 4, by taking
effects of a tiny but nonzero positive cosmological constant into
account, we show that the theoretic threshold is above the
energies of all the observed UHECR events. Similar discussion is
made for the TeV-$\gamma$ in Section 5. In the end, we present
conclusions and remarks.

\section{Classical kinematics}

The dS spacetime can be realized as a four dimensional
pseudo-sphere imbedded in the five dimensional flat space

\begin{equation}\label{xi}
\begin{array}{l}
\displaystyle
\left(\xi^0)^2-(\xi^1)^2-(\xi^2)^2-(\xi^3)^2-(\xi^5\right)^2=-\frac{1}{\lambda}~,\\[5mm]
ds^2=\left(d\xi^0)^2-(d\xi^1)^2-(d\xi^2)^2-(d\xi^3)^2-(d\xi^5\right)^2~,
\end{array}
\end{equation}
where $\lambda$ is the Riemannian curvature of the dS spacetime.
It is obvious that the above equations are invariant under the
action of the de Sitter group $SO(1,~4)$. The coordinates
${\xi^{\mu}}$ ($\mu=0,~1,~2,~3,~5$) are related to the Beltrami
coordinates $x^i$ $(i=0,~1,~2,~3)$ through the relation
\begin{equation}\label{xi-x}
x^i=\frac{\xi^i}{\sqrt{\lambda}\xi^5}~,~~~~~(\xi^5\not=0)~.
\end{equation}
Then, in the Beltrami coordinate, we can write the dS spacetime
as\cite{wulixuebao-1,1979-6}

\begin{equation}\label{sigma}
\begin{array}{l}
\sigma\equiv\sigma(x,~x)=1-\lambda\eta_{ij}x^ix^j>0~,\\[3mm]
\displaystyle ds^2=\left(\frac{\eta_{ij}}{\sigma}+\frac{\lambda\eta_{ir}\eta_{js}x^rx^s}
{\sigma^2}\right)dx^idx^j~,
\end{array}
\end{equation}
where $\eta_{ij}={\rm diag}(1,~-1,~-1,~-1)$ is the Minkowski
metric. Transformations of $x^i$ acted by $SO(1,~4)$ can be
expressed as follows,

\begin{equation}
\begin{array}{l} \label{x-x-tilde}
x^i\rightarrow\tilde{x}^i=\sigma(b,~b)^{1\over2}\sigma(b,~x)^{-1}(x^j-b^j)D^i_j~,\\
[0.5cm]
D^i_j=L^i_j+\lambda\left(\sigma(b,~b)+\sigma(b,~b)^{1\over2}\right)^{-1}\eta_{kl}b^lb^iL^k_j~,\\
[0.5cm] L\equiv(L^i_j)~~\in SO(3,~1)~,\\
[0.5cm]\sigma(b,~b)>0~,
\end{array}
\end{equation}
where $(b^i)$ is an arbitrary point in the Beltrami dS spacetime.
We can define the five dimensional angular momentum $M^{\mu\nu}$
of a free particle with mass $m_0$ as the form
\begin{equation}\label{angular}
M^{\mu\nu}=m_0\left(\xi^{\mu}\frac{d\xi^{\nu}}{ds}-\xi^{\nu}\frac
{d\xi^{\mu}}{ds}\right)~,
\end{equation}
where $s$ is a parameter along the geodesic. In the dS spacetime,
there is no translation invariance and so that one can not
introduce a momentum vector. However, it should be noticed that,
at least somehow, we may define a counterpart of the momentum $P$
for a free particle in the dS spacetime
\begin{equation}
\label{momentum}
P^i\equiv\sqrt{\lambda}M^{5i}=m\sigma^{-1}\frac{dx^i}{ds}~,~~~~
(i=0,~1,~2,~3)~.
\end{equation}
In the same manner, the counterparts of the four dimensional
angular momentum $J^{ij}$ can be assigned as
\begin{equation}
\label{angular-mn} J^{ij}\equiv M^{ij} =
x^iP^j-x^jP^i=m\sigma^{-1}\left(x^i\frac{dx^j}{ds}-x^j\frac{dx^i}{ds}\right)~.
\end{equation}
It is not difficult to show that, along the geodesics, the above
defined momentum and angular momentum are invariant,
\begin{equation}\label{conserve-P}
 \frac{dP^i}{ds}=0~, ~~~~~\frac{dJ^{ij}}{ds}=0~.
\end{equation}
In fact, these are the equations of geodesic in the dS spacetime.
From Eq. (\ref{conserve-P}), we get, for a free particle,
\begin{equation}
\frac{dx^\alpha}{dx^0}={\rm constant}~,~~~~~~(\alpha=1,~2,~3)~.
\end{equation}
The fact tells us that the Beltrami coordinate can be considered
to be an inertial frame\cite{kexuetongbao} as the Cartesian
coordinate in the Minkowski spacetime.  We notice  that there is a
subgroup $SO(4)$ of the de Sitter one $SO(1,4)$, which consists of
spatial  transformations among $x^\alpha$. It is easy to show that
$\xi^0(\equiv\sigma(x,x)^{-1/2}x^0)$ is invariant under the
spatial transformations. Thus, we can say  two spacelike events
are simultaneous if they satisfy
\begin{equation}
\label{simultaneous} \sigma(x,~x)^{-\frac{1}{2}}x^0=\xi^0={\rm
constant}~.
\end{equation}
Therefore, it is convenient to discuss physics of the dS spacetime
in the coordinate $(\xi^0,~x^\alpha)$. In this coordinate, the
metric can be rewritten into the form

\begin{equation}
ds^2=\frac{d\xi^0d\xi^0}{1+\lambda\xi^0\xi^0}-(1+\lambda\xi^0\xi^0)
\left[\frac{d\rho^2}{(1+\lambda\rho^2)^2}
+\frac{\rho^2}{1+\lambda\rho^2}d\Omega^2\right]~,
\end{equation}
where $\rho^2\equiv\Sigma{x^{\alpha}x^{\alpha}}$ and $d\Omega^2$
denotes the metric on 2-dimensional sphere $S^2$.\\
If a proper time $\tau$ is introduced as
\begin{equation}
\tau\equiv
\frac{1}{\sqrt{\lambda}}\sinh^{-1}(\sqrt{\lambda}\xi^0)~,
\end{equation}
one would get a Friedman-Robertson-Walker like metric,
\begin{equation}
ds^2=d\tau ^2 -\cosh^2(\sqrt{\lambda}\tau)
\left[\frac{d\rho^2}{(1+\lambda\rho^2)^2}
+\frac{\rho^2}{1+\lambda\rho^2}d\Omega^2\right]~.
\end{equation}

In terms of the five dimensional angular momentum $M^{\mu\nu}$, we
can construct an invariant under the de Sitter transformations for
a free particle\cite{wulixuebao-2},
\begin{equation}
\begin{array}{c}
\label{casimir}
m^2_0=\displaystyle\frac{\lambda}{2}M^{\mu\nu}M_{\mu\nu}=E^2-{\bf
P}^2+\frac{\lambda}{2} J^{ij}J_{ij}~,\\ [0.5cm]
 E=P^0~,~~~~{\bf
P}=(P^1,~P^2,~P^3)~.
\end{array}
\end{equation}
However it should be noticed that the $P^i$ defined in this way is
not a four dimensional vector but the component of the five
dimensional angular momentum $M^{\mu\nu}$.

\section{Quantum kinematics}

It is natural to realize the five dimensional angular momentum
$M_{\mu\nu}$ as infinitesimal generators of the de Sitter group
$SO(1,~4)~$
\begin{equation}
\label{L-operator}
M^{\mu\nu}=-i\left(\xi^{\mu}\frac{\partial}{\partial\xi_{\nu}}-\xi^{\nu}
\frac{\partial}{\partial\xi_{\mu}}\right)~.
\end{equation}
The de Sitter invariant, or the Casimir operator can be used to
express the one-particle states in the dS spacetime
\begin{equation}
\label{invariant}
\left(\frac{\lambda}{2}M^{\mu\nu}M_{\mu\nu}-m^2_0\right)\Phi(\xi^0,x^{\alpha})=0~,
\end{equation}
where the $\Phi(\xi^0,x^{\alpha})$ denotes a scalar field or a
component of vector fields with a given spin $s$.

In the coordinates $(\xi^0,~x^{\alpha})$~, we can rewrite the
Casimir operator as the following form
\begin{eqnarray}
\frac{\lambda}{2}M^{\mu\nu}M_{\mu\nu}&=&-\left(
1+\lambda\xi^0\xi^0
\right)\partial_{\xi^0}^2-4\lambda\xi^0\partial_{\xi^0}  \nonumber \\
&&+\left( 1+\lambda\xi^0\xi^0 \right)^{-1}\left( 1+\lambda\rho^2
\right)^2\left[ \partial^2_\rho+2\rho^{-1}\partial_\rho \right] \\
&&+\left( 1+\lambda\xi^0\xi^0 \right)^{-1}\left( 1+\lambda\rho^2
\right)\rho^{-2} \left(\partial^2_{\bf u}-s(s+1)\right) ~,
\nonumber
\end{eqnarray}
where ${\bf u}{\bf u}^{\prime}=1$ and $\partial^2_{\bf u}$ denotes
the Laplace operator on $S^2$.

To solve the equation of motion, we write the field
$\Phi(\xi^0,x^{\alpha})$ into the form
\[
\Phi(\xi^0,\rho,{\bf u})=T(\xi^0)U(\rho)Y_{lm}(\bf u)~.
\]
Thus, we transform the equation of motion into
\cite{ds-paper-3,wulixuebao-2},
\begin{eqnarray}
&&
\left[(1+\lambda\xi^0\xi^0)^2\partial^2_{\xi^0}+4\lambda\xi^0(1+\lambda\xi^0\xi^0)
\partial_{\xi^0}+m_0^2(1+\lambda\xi^0\xi^0)+(\varepsilon^2-m_0^2)\right]T(\xi^0)=0,
\nonumber \\[0.5cm]
&& \left[\partial^2_{\rho}+
\displaystyle\frac{2}{\rho}\partial_\rho-\left[\frac{m_0^2-\varepsilon^2}
{(1+\lambda\rho^2)^2}+\frac{l(l+1)+s(s+1)}{\rho^2(1+\lambda\rho^2)}\right]
\right]U(\rho)=0, \nonumber\\[0.5cm]
&& \left[\partial^2_{\bf u}+l(l+1)\right]Y_{lm}({\bf u})=0,
\end{eqnarray}
where $Y_{lm}({\bf u})$ is the spherical harmonic function and
$\varepsilon$ is a constant.

Solutions of timelike part of the field are of the forms

\begin{equation}
T(\xi^0)\sim\left(1+\lambda\xi^0\xi^0\right)^{-1/2} \cdot\left\{
\begin{array}{l}
P_{\nu}^{\mu}(i\sqrt{\lambda}\xi^0)~, \\
\\
Q_{\nu}^{\mu}(i\sqrt{\lambda}\xi^0) ~,%
\end{array}
\right.
\end{equation}
where $\mu,~\nu~$ satisfy
\newline
\begin{equation}
\begin{array}{l}
\nu(\nu+1)=2-\lambda^{-1}m_0^2 ~, \\
\mu^2= 1+\lambda^{-1}(\varepsilon^2-m_0^2) ~.%
\end{array}%
\end{equation}
For the radial equation of the field, we can write the solutions
as the form
\begin{equation}
U(\rho)\sim\rho^l(1+\lambda\rho^2)^{k/2}F\left(\frac{1}{2}(l+k+s+1),~\frac{1}{2}
(l+k+s),~l+s+\frac{3}{2};~-\lambda\rho^2\right)~,
\end{equation}
where $k$ denotes the radial quantum number
$$k^2-2k-\lambda^{-1}(\varepsilon^2-m_0^2)=0~.$$
To be normalizable, the hypergeometric function in the radial part
of the wavefunction has to break off, leading to the quantum
condition

\begin{equation}
\frac{l+k+s}{2}=-n~,~~~~~(n\in \mathbf{N})~.
\end{equation}
Then, we obtain the dispersion relation for a free particle

\begin{equation}
\label{dispersion2}
E^2=m_0^2+\varepsilon'^2+\lambda(2n+l+s)(2n+l+s+2)~.
\end{equation}

\section{UHECR threshold anomaly}

Until today, we have observed\cite{data1}-\cite{data5} hundreds of
events with energies above $10^{19}$eV and about 20 events above
$10^{20}$eV, which are above the GZK threshold. In principle,
photopion production with the cosmic microwave background
radiation photons should decrease the energies of these protons to
the level below the corresponding threshold. In this section, we
discuss the UHECR threshold anomaly in the covariant framework of
kinematics in dS spacetime set up in the preceding sections.

We consider the head-on collision between a soft photon of energy
$\epsilon$, momentum ${\bf q}$ and a high energy particle $m_1$ of
energy $E_1$, momentum ${\bf p}_1$, which leads to the production
of two particles $m_2,~m_3$ with energies $E_2$, $E_3$ and
momentums ${\bf p}_2$, ${\bf p}_3$, respectively. From the energy
and momentum conservation laws, we have

\begin{equation}\label{threshold1}
\begin{array}{c}
E_1+\epsilon=E_2+E_3~, \\[0.4cm]
p_1-q=p_2+p_3~.
\end{array}
\end{equation}
In the C.M. frame, $m_2$ and $m_3$ are at rest at threshold, so
they have the same velocity in the lab frame and there exists the
following relation

\begin{equation}
\label{p3p2} \frac{p_2}{p_3}=\frac{m_2}{m_3}~.
\end{equation}
It is convenient to use the approximated formulae of dispersion
relations (\ref{dispersion2}) for the soft photons and the ultra
high energy particles
\begin{eqnarray}\label{dispersion-app}
&&\epsilon^2=q^2+\lambda_{\gamma}^{*}~,\\[0.4cm]
&&E_i=\sqrt{m^2_i+p^2_i+\lambda_i^{*}}\simeq
p_i+\frac{m^2_i}{2p_i}+\frac{\lambda^{*}_i}{2p_i}~,~~(i=1,~2,~3)~.
\end{eqnarray}
where
$~\lambda_{\gamma}^{*}\equiv\lambda(l_\gamma+2n_\gamma+1)(l_\gamma+2n_\gamma+3)
\approx\lambda {l_{\gamma}}(l_{\gamma}+1)~$ and
$~\lambda_{i}^{*}\equiv\lambda(l_i+2n_i+s_i)(l_i+2n_i+s_i+2)\approx\lambda
{l_{i}}(l_{i}+1)~$ with the conjecture that $~l \gg n,~s~$.

The obtained threshold can be expressed as the form

\begin{equation}
\label{threshold0}\displaystyle E_{{\rm
~th},~\lambda}\simeq\frac{(m_{2}+m_{3})^2-m_{1}^2+\lambda_{2}^{*}
\left(1+\frac{m_3}{m_2}\right)+\lambda_{3}^{*}\left(1+\frac{m_2}{m_3}\right)-
\lambda_{1}^{*}}{2\left(\epsilon+\sqrt{\epsilon^2-\lambda_{\gamma}^{*}}\right)}~.
\end{equation}
The usual GZK threshold could be recovered when the parameter
$\lambda^*$, which is dependent on the cosmological constant, runs
to zero.

The conservation law of the angular momentum gives a constraint on
the parameters $\lambda^*$,
$$
\lambda^{*}_{1} + \lambda^{*}_{\gamma} + 2\lambda{\bf
L}_1\cdot{\bf L}_{\gamma} = \lambda^{*}_{2} + \lambda^{*}_{3} +
2\lambda{\bf L}_2\cdot{\bf L}_3~.
$$
Making use of the relation, we can rewrite the $\lambda^*$
dependent terms of the threshold as the following

\begin{equation}
\frac{\lambda^{*}_{2}\frac{m_3}{m_2}+\lambda^{*}_{3}\frac{m_2}{m_3}+\lambda^{*}_{\gamma}
+2\lambda{\bf L}_1\cdot{\bf L}_{\gamma}-2\lambda{\bf L}_2\cdot{\bf
L}_3}
{2\left(\epsilon+\sqrt{\epsilon^2-\lambda_{\gamma}^{*}}\right)}~.
\end{equation}
If $\lambda^{*}_2$ and $\lambda^{*}_3$  take value of the same
order with $\lambda^{*}_{\gamma}$ (less than the square of energy
of a soft photon), the $\lambda^*$ dependent terms can be
omitted\cite{0307439}.  We will investigate the case of
$\lambda^{*}_2+\lambda^{*}_3 \gg \lambda^{*}_{\gamma}~$, and the
threshold $(\ref{threshold0})$ is of the form

\begin{equation}
\label{threshold}\displaystyle E_{{\rm
~th},~\lambda}\simeq\frac{(m_{2}+m_{3})^2-m_{1}^2+
\lambda^{*}_{2}\frac{m_3}{m_2}+\lambda^{*}_{3}\frac{m_2}{m_3}-
2\lambda{\bf L}_2\cdot{\bf L}_3}{2\left(\epsilon+
\sqrt{\epsilon^2-\lambda_{\gamma}^{*}}\right)}~.
\end{equation}

\vspace*{0.5cm}

Now, we can study the photopion production processes of the UHECR
interaction with the CMBR
$$p+\gamma\rightarrow p+\pi~.$$
The corresponding threshold for this process is given by

\begin{equation}
\label{UHECR-threshold}\displaystyle E^{\rm UHECR}_{{\rm
~th},~\lambda}\simeq\frac{(m_{N}+m_{\pi})^2-m_{N}^2+
\lambda^{*}_{N}\frac{m_{\pi}}{m_N}+\lambda^{*}_{\pi}\frac{m_N}{m_{\pi}}-
2\lambda{\bf L}_{N}\cdot{\bf L}_{\pi}}{2\left(\epsilon+
\sqrt{\epsilon^2-\lambda_{\gamma}^{*}}\right)}~.
\end{equation}
To show the behavior of the threshold in the $\lambda^*$-parameter
space clearly, we should discuss some limit cases in detail.

In the case that the out-going nucleon has zero angular momentum,
the threshold $(\ref{UHECR-threshold})$  reduces  as

\begin{equation}
\label{UHECR-pi}\displaystyle E^{\rm UHECR}_{{\rm
~th},~\lambda,\pi}\simeq\frac{(m_{N}+m_{\pi})^2-m_{N}^2+
\lambda^{*}_{\pi}\frac{m_N}{m_{\pi}}}{2\left(\epsilon+
\sqrt{\epsilon^2-\lambda_{\gamma}^{*}}\right)}~.
\end{equation}
We give a plot for the dependence of the threshold $E^{\rm
UHECR}_{{\rm th},~\lambda,\pi}$ on the cosmological constant and
angular momentum (the in-going photon and out-going pion) in {\bf
FIG.1}.
\begin{center}
  \includegraphics[height=95mm,width=100mm]{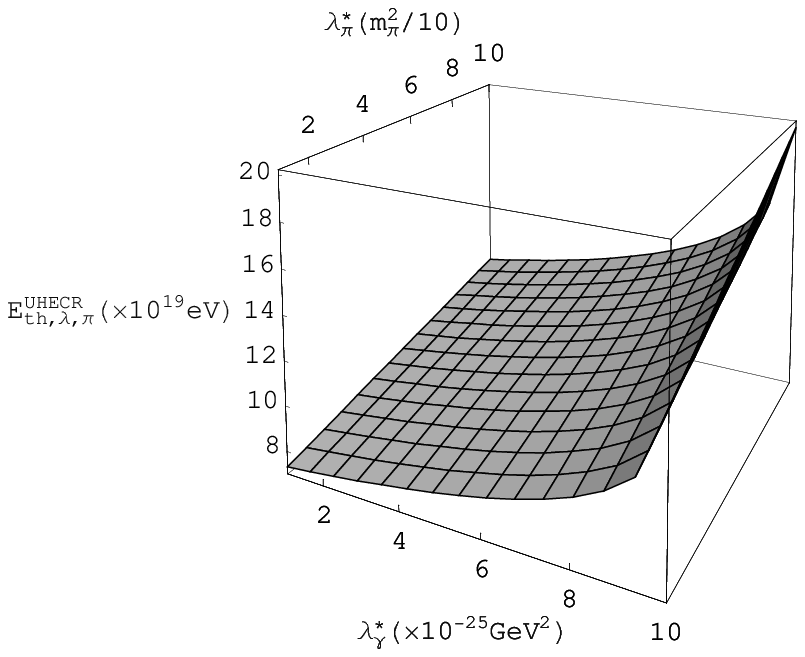}
\end{center}
{\bf FIG.1}~~~~ {\scriptsize The cosmological constant and angular
momentum (of in-going soft photon and out-going pion) dependence
of the threshold $E^{\rm UHECR}_{{\rm th},~\lambda,\pi}$ in the
interaction between the UHECR protons and the CMBR photons
($\lambda^{*}_{\pi}$ in unit of $m_{\pi}^2/10$)~.}

In the case that the out-going pion has zero angular momentum, the
UHECR threshold takes the form

\begin{equation}
\label{UHECR-pn}\displaystyle E^{\rm UHECR}_{{\rm
~th},~\lambda,N}\simeq\frac{(m_{N}+m_{\pi})^2-m_{N}^2+
\lambda^{*}_{N}\frac{m_{\pi}}{m_{N}}}{2\left(\epsilon+
\sqrt{\epsilon^2-\lambda_{\gamma}^{*}}\right)}~.
\end{equation}
We give a plot for the dependence of the threshold $E^{\rm
UHECR}_{{\rm th},~\lambda,N}$ on the cosmological constant and
angular momentum (the in-going soft photon and out-going nucleon)
in {\bf FIG.2}.
\begin{center}
  \includegraphics[height=95mm,width=100mm]{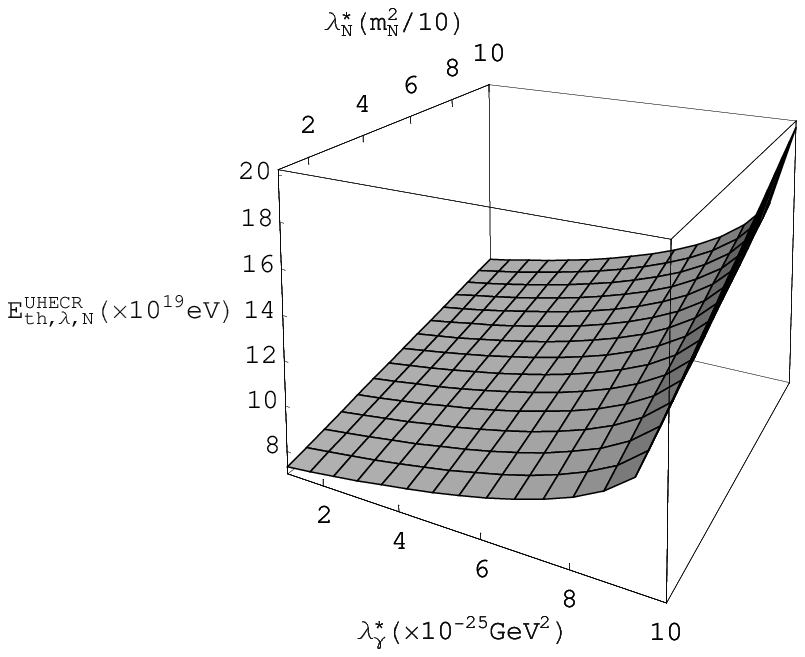}
\end{center}
{\bf FIG.2}~~~~ {\scriptsize The cosmological constant and angular
momentum (of the in-going soft photon and out-going nucleon)
dependence of the threshold $E^{\rm UHECR}_{{\rm th},~\lambda,N}$
in the interaction between the UHECR protons and the CMBR photons
($\lambda^{*}_{N}$ in unit of $m_{N}^2/10$)~.}

Finally, if the out-going pion and nucleon has the same angular
momentum, the UHECR threshold can be expressed as the following
form

\begin{equation}
\label{UHECR-pnpi}\displaystyle E^{\rm UHECR}_{{\rm
~th},\lambda,N\pi}\simeq\frac{(m_{N}+m_{\pi})^2-m_{N}^2+
\lambda^{*}_{N}(\frac{m_{N}}{m_{\pi}}+\frac{m_{\pi}}{m_{N}}-2)}{2\left(\epsilon+
\sqrt{\epsilon^2-\lambda_{\gamma}^{*}}\right)}~.
\end{equation}
We give a plot for the dependence of the threshold $E^{\rm
UHECR}_{{\rm th},~\lambda,N\pi}$ on the cosmological constant and
angular momentum (the out-going nucleon and pion)  in {\bf FIG.3}.

From the above discussion, we know that  a tiny but nonzero
cosmological constant may provide indeed sufficient corrections to
the primary predicted threshold\cite{GZK}. For the observed
cosmological constant (which is around the level of $10^{-85}{\rm
GeV}^2$), if the CMBR possesses a quantum number $l_{\gamma}$ of
the order of $10^{30}$, the threshold will be above the energies
of all those observed UHECR particles.  The predicted threshold
should be upgraded to a more reasonable level. Now we can say that
a possible origin of the cosmic ray threshold anomaly has been
got. It is the cosmological constant that increases the GZK
cut-off to a level above the observed UHECR events.
\begin{center}
  \includegraphics[height=95mm,width=100mm]{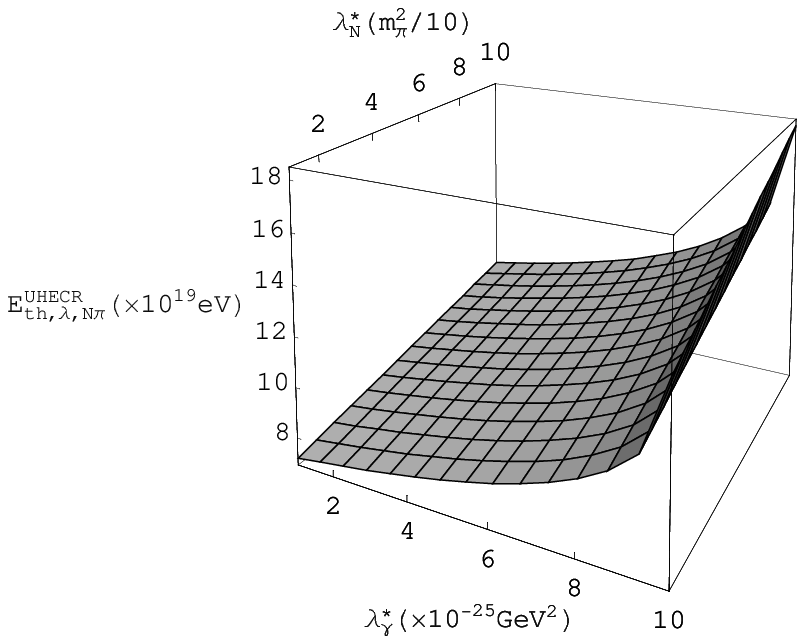}
\end{center}
{\bf FIG.3}~~~~ {\scriptsize The cosmological constant and angular
momentum (of the out-going nucleon and pion) dependence of the
threshold $E^{\rm UHECR}_{{\rm th},~\lambda,N\pi}$ in the
interaction between the UHECR protons and the CMBR photons
($\lambda^{*}_{N}=\lambda^{*}_{\pi}$ in unit of $m_{\pi}^2/10$)~.}

\section{TeV-$\gamma$ threshold anomaly}

The second paradox for the cosmic ray comes from the fact that
experiments detected 20TeV photons from Mrk 501 (a BL Lac object
at a distance of 150Mpc). Similar to the UHECR case, due to the
interaction with the IR background photons, the 20TeV photons
should have disappeared before arrival at the ground-based
detections. Parallel to the analysis of UHECR in the previous
section, we discuss the TeV-$\gamma$ threshold anomaly in the
following.

For the pair production process
$$\gamma+\gamma\rightarrow e^++e^-~,$$
we obtain a threshold by taking $m_1$ in the equation
$(\ref{threshold})$ to be zero,

\begin{equation}
\label{Tev-threshold} E^{\gamma}_{{\rm
~th},~\lambda}\simeq\frac{4m_e^2+\lambda^{*}_{e^{+}}+
\lambda^{*}_{e^{-}}-2\lambda{\bf L}_{e^+} \cdot{\bf
L}_{e^-}}{2(\epsilon+\sqrt{\epsilon^2-\lambda^{*}_{\gamma}})}~.
\end{equation}
If one of the out-going particles possesses zero angular momentum,
the corresponding threshold is reduced to the form

\begin{equation}
\label{Tev-threshold} E^{\gamma}_{{\rm
th},\lambda,e}\simeq\frac{2m_e^2+\lambda^{*}_{e}/2}
{\epsilon+\sqrt{\epsilon^2-\lambda^{*}_{\gamma}}}~.
\end{equation}
We present a plot of the dependence of the threshold
$E^{\gamma}_{{\rm th},\lambda,e}$ on the cosmological constant and
angular momentum (the out-going electron or positron)  in {\bf
FIG.4}.
\begin{center}
  \includegraphics[height=95mm,width=100mm]{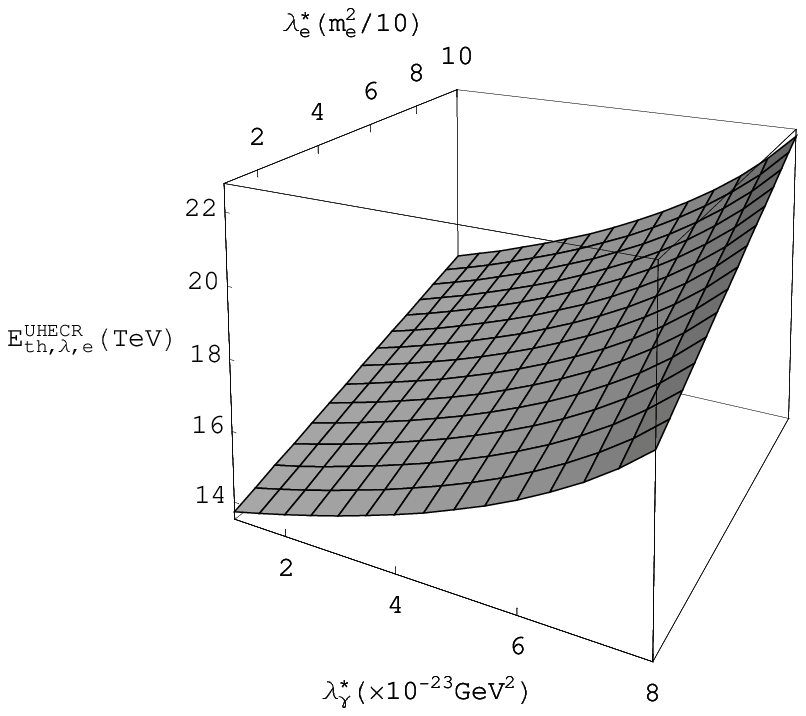}
\end{center}
{\bf FIG.4}~~~~ {\scriptsize The cosmological constant and angular
momentum ( of the in-going soft photon and out-going electron)
 dependence of the threshold $E^{\gamma}_{{\rm th},\lambda,e}$ in
the interaction between the TeV-$\gamma$ and the IR background
photons ($\lambda^{*}_{e}$ in unit of $m_{e}^2/10$)~.}

From the above plot, we  see that the threshold $E^{\gamma}_{{\rm
th},\lambda}$ is also quite sensitive to the small positive
cosmological constant.  By taking into account the corrections
which are dependent on the cosmological constant and  angular
momentum of  interacting particles, we get a new predicted
threshold. And then the threshold anomaly disappears.

\section{Conclusions and remarks}
In this paper, we discussed kinematics in the dS spacetime. The
kinematic invariance group in the dS spacetime is $SO(1,4)$
instead of the Poincar\'{e} one in the Minkowski sapcetime.
Kinematics, which is based on the de Sitter group $SO(1,4)$, was
set up formally.  It should be noticed that the dispersion
relation for a free particle in the dS spacetime is related with
degrees of freedom of angular momentum and spin. With the help of
this deformed dispersion relation, and thanks to the positive
cosmological constant, a possible origin of the threshold
anomalies was proposed naturally.

We would like to point out that, if the cosmological constant can
vary notably in different period of the Universe (at least in some
scenarios it is), we may get a higher threshold for the UHECR
propagating in the early Universe. And this could be a good news
for the ``top-down" scenario of the UHECR.
\newline
\newline
{\large\bf Acknowledgement:}\\
We would like to thank Prof. H. Y. Guo and C. J. Zhu for useful
discussion. The work was supported partly by the Natural Science
Foundation of China. One of us (C.B.G.) is supported by grants
through the ICTS (USTC) from the Chinese Academy of Sciences.

\end{document}